\documentclass{ws-mpla}
\usepackage[super]{cite}
\usepackage{graphicx}

\newcommand{\be}{\begin{eqnarray}}
\newcommand{\ee}{\end{eqnarray}}
\newcommand{\n}{\nonumber}
\newcommand{\eps}{\epsilon}
\newcommand{\lam}{\lambda}
\newcommand{\ba}{\begin{array}}
\newcommand{\ea}{\end{array}}

\begin{document}

\markboth{C.-L. Ho and P. Roy}
{Dirac equation with complex potentials}

\catchline{}{}{}{}{}

\title{Dirac equation with complex potentials}

\author{\footnotesize C.-L. Ho}

\address{Department of Physics, Tamkang University, Tamsui
25137, Taiwan\\
hcl@mail.tku.edu.tw}

\author{P. Roy}

\address{Physics and Applied Mathematics Unit, Indian Statistical Institute, Kolkata 700108,
India\\
pinaki@isical.ac.in}

\maketitle

\pub{Received (Day Month Year)}{Revised (Day Month Year)}

\begin{abstract}
We study $(2+1)$ dimensional Dirac equation with complex scalar and Lorentz scalar potentials. It is shown that the Dirac equation admits exact analytical solutions with real eigenvalues for certain complex potentials while for another class of potentials zero energy solutions can be obtained analytically.  For the scalar potential cases, it has also been shown that the {\it effective} Schr\"odinger-like equations resulting from decoupling the spinor components can be interpreted as exactly solvable energy dependent Schr\"odinger equations.

\keywords{Dirac equation; Complex potentials; Energy-dependent potentials}
\end{abstract}

\ccode{PACS Nos.: 03.65.-w, 03.65.Pm, 03.65.Ge}

\section{Introduction}
Since the seminal work by Bender and Boettcher \cite{b1} there have been numerous investigation on $\cal{PT}$ symmetric as well as pseudo Hermitian quantum mechanics \cite{mostafa}. Initially most of the papers dealt with non Hermitian non relativistic quantum mechanics. Later the concept was extended to relativistic quantum mechanics as well \cite{rel1,rel2,rel3,rel4,rel5}. In particular non Hermitian relativistic quantum mechanics can be realized in the field of optics \cite{longhi}. It is generally believed that $\cal{PT}$ symmetry ensures that the spectrum is a real one unless $\cal{PT}$ symmetry is spontaneously broken \cite{zafar}. However in a recent paper it has been shown that non $\cal{PT}$ symmetric Dirac equation too can admit real eigenvalues \cite{al}. Here our objective is to demonstrate that Dirac equation with certain types of complex scalar potentials or electric fields can also admit real eigenvalues. In particular we shall examine two types of potentials - in one type of potentials all the energy eigenvalues can be found analytically while for the second type of potentials exact zero energy states can be found. It will also be seen that {\it effective} Schr\"odinger-like equations for the spinor components can be interpreted as exactly solvable energy dependent Schr\"odinger equations \cite{GGPS,YLL}. 
\section{The model}
The massless stationary Dirac-Weyl equation in the presence of
a one-dimensional scalar potential $V\left(x\right)$ is given by \cite{dirac}
\begin{equation}\label{dirac1}
\left[c\left(\sigma_{x}\hat{p}_x+\sigma_y\hat{p}_y\right)+V\left(x\right)\right]\Psi=E\Psi,
\end{equation}
where $\sigma_{x,y}$ are the Pauli spin matrices, $\Psi$ is a two component spinor
and $c$ is the velocity of light.

Here we would like to consider the possibility of a real spectrum of Eq.(\ref{dirac1}) with a complex potential $V(x)$.
As $V(x)$ depends only on $x$, one can assume the  two-component
Dirac wavefunction to be of the form:
\begin{equation}
\Psi (x)=e^{ik_y y}\left(\begin{array}{c}
\psi_{A}\left(x\right)\\
\psi_{B}\left(x\right)
\end{array}\right).
\end{equation}
The Dirac equation is then reduced to the following pair of coupled equations
\be
&&\left(U\left(x\right)-\epsilon\right)\psi_{A}-i\left(\frac{d}{dx}+k_{y}\right)\psi_{B}=0,
\label{dirac_1}\\
&&\left(U\left(x\right)-\epsilon\right)\psi_{B}-i\left(\frac{d}{dx}-k_{y}\right)\psi_{A}=0.
\label{dirac_2}
\ee
where $U=V/c$ and $\epsilon=E/c$. 

It is interesting to note that Eqs.~(\ref{dirac_1}) and (\ref{dirac_2}) are invariant under the following transformations:
\[
k_y\to -k_y,~~ \psi_A\leftrightarrow \psi_B.
\]
This means that if the spinor $\psi=e^{ik_yy}(\psi_A,\psi_B)^t$ is a solution for $k_y$, then $\psi=e^{-ik_yy}(\psi_B,\psi_A)^t$ is a solution for $-k_y$ (here ``$t$" means transpose).
That is, eignestates with opposite signs of $k_y$ are spin-flipped. 

For $k_y=0$, Eqs.~(\ref{dirac_1}) and (\ref{dirac_2}) are invariant under the changes $\psi_A\leftrightarrow \pm \psi_B$, and the respective solutions are
\be
\psi(x)=\psi_A(x)\left(\ba{c}1\\\pm 1\ea\right),~~\psi_A(x)=\exp\left(\mp i \int^x (U(x)-\eps)\, dx\right).
\label{k0}
\ee
It is worthy to note that in this case, the wave function $\psi(x)$ is normalizable only in a finite domain  if $U(x)$ is real.  For $\psi(x)$ to be normalizable on a half-line or the whole line, $U(x)$ need to be complex.

For $k_y\neq 0$, it is convenient to make the substitution $\psi_{A}=\left(\psi_{+}+\psi_{-}\right)/2$ and $\psi_{B}=\left(\psi_{-}-\psi_{+}\right)/2$.
Eqs.~(\ref{dirac_1}) and (\ref{dirac_2}) then become
\begin{eqnarray}
&&\left(U\left(x\right)-\epsilon-i\frac{d}{dx}\right)\psi_- +ik_{y}\psi_+=0,\label{dirac2}\\
&&\left(U\left(x\right)-\epsilon+i\frac{d}{dx}\right)\psi_+ -ik_{y}\psi_-=0.\label{dirac3}
\end{eqnarray}
These coupled equations can then be further reduced to a pair of decoupled Schr\"odinger-like  equation in $\psi_+$ $\left(\psi_-\right)$
\begin{equation}
\left(-\frac{d^2}{dx^2}+ U_\mp(x)\right)\psi_{\mp}=0,
\label{decoupled}
\end{equation}
with \emph{energy-dependent} effective potentials
\begin{equation}\label{upm}
U_\mp(x)=-\left(U(x)-\epsilon\right)^{2}\mp i\frac{dU(x)}{dx}+k_{y}^{2}.
\end{equation}

In the following discussion  we shall concentrate on the upper component $\psi_-$. 
The wave function $\psi_+$ can be obtained from Eq.~(\ref{dirac2}).

\section{An Exactly solvable potential}

In this section we consider an example whose spectra are real and exactly solvable.

Let us take
\[
U(x)=iV_0\cot  x, ~~~V_0>0,~~ 0<x<\pi.
\]
The potential $U_-(x)$ corresponding to the component $\psi_-$ is
\begin{equation}
U_-(x)=V_0\left(V_0-1\right)\,{\rm cosec}^2 x + 2i\epsilon V_0 \cot x -\left(\epsilon^2-k_y^2+V_0^2\right).
\label{U-RM}
\end{equation}
This effective potential has the form of the Rosen-Morse I potential \cite{khare}
 \begin{equation}
V(x)=A\left(A-1\right)\,{\rm cosec}^2 x+2B \cot x 
\end{equation}
whose energy eigenvalues are given by
\begin{equation}
E_n=\left[\left(A+n\right)^2-\frac{B^2}{(A+n)^2}\right], ~~ n=0,1,2,\ldots\\
\end{equation}
The main difference is that here the potential (\ref{U-RM}) is {\it energy-dependent}.
Comparing the corresponding terms lead to
\begin{eqnarray}
&&~~~~~~~~~ A=V_0,~~B=i\epsilon_nV_0,\nonumber\\
&&\epsilon_n^2-V_0^2-k_y^2=-\left(V_0+n\right)^2 + \frac{\epsilon_n^2 V_0^2}{(V_0+n)^2}.
\end{eqnarray}
We have included the subscript ``$n$" to indicate the $n$-th eigenenergy. 
The eigenenrgies are thus given by
\begin{equation}
\epsilon_n^2=\frac{1}{\left[1-\frac{V_0^2}{(V_0+n)^2}\right]}\,\left(n^2+2V_0 n  + k_y^2\right).
 \end{equation}
The corresponding wavefunction is \cite{khare}
\begin{equation}
\psi_-\sim (y^2-1)^{-(s+n)/2}\, P_n^{(-s-n+ia,-s-n-ia)}(y),
\end{equation}
where
\[
y=i\cot x,~~s=V_0,~~a=i\frac{\epsilon_n V_0}{s+n},
\]
and $P_n^{(\alpha,\beta)}(y)$ are the classical Jacobi polynomials.

\vspace{.5cm}


\section{Solvable zero energy states}
In this section we shall present several examples of complex potentials for which zero energy $(\epsilon=0$) states can be found out analytically. 

\subsection{Example 1}

First let us consider a very simple potential of the form
\be
U(x)=(x-i\mu)^2,~~~~\mu:~{\rm real}.
\ee
Then the effective potentials are given by
\be
U_\pm(x)=-(x-i\mu)^4\pm 4i(x-i\mu)^3+k_y^2.
\ee
One may verify directly that
\be
\psi_\pm(x)=e^{\pm i(x^3/3-\mu^2x-i\mu x^2)}
\ee
satisfy the eigenvalue equations
\be
\left[-\frac{d^2}{dx^2}+U_\pm(x)\right]\psi_\pm=0,
\label{EqU5}
\ee
when $k_y=0$. Thus in this case the zero energy solution is non degenerate.

However, $\psi_+$ does not approach zero as $x\to \pm\infty$.  So a viable solution for the system is to take the trivial solution
namely, $\psi_+=0$ for the potential $U_+$.
This leads to $\psi_A=\psi_B=\psi_-$, which is consistent with Eq. (\ref{k0})  with $\eps=k_y=0$.

\subsection{Example 2}

Here we shall consider a complex potential $U(x)$ such that the effective potential in the decoupled equations for the components would be a complex hyperbolic secant (or its variants) type potential. In the present case we take 
\be\label{ex4}
U(x)= -i\mu\tanh x + \lambda\, {\rm sech} \,x, ~~\mu, \lambda:~{\rm real}.\ee
Then from (\ref{upm}) we find
\be\label{U3}
U_-(x)=\mu^2-(\lambda^2+\mu^2+\mu)\, {\rm sech}^2 x +  i\lambda (2\mu+1) {\rm sech}\, x \tanh x+k_y^2.
\ee
The potentials in (\ref{U3}) can be identified with complexified or more precisely $\cal{PT}$ symmetrized Scarf II potential of the form
\be\label{scarf}
V_{Scarf II}(x)=A^2 -(B^2+A^2+A){\rm sech}^2x+iB(2A+1)\,{\rm  sech} \,x\,\tanh\,x.
\ee
The potential (\ref{scarf}) is exactly solvable with energy eigenvalues ($E_n$) and wavefunctions ($\phi_n$) given by \cite{khare}
\be\label{scarfsol}
E_n &=&A^2-(A-n)^2,~~~~y=i\sinh\,x,~~~~n=0,1,2....<[A-1],\\
\phi_n&=&i^n(1+y^2)^{-A/2}~e^{-B\,\tan^{-1}y}P_n^{(iB-A-1/2,-iB-A-1/2)}(y),\label{phi_4}
\ee
where $P_n^{(a,b)}(x)$ denotes Jacobi polynomials. Now comparing $U_-(x)$ with (\ref{scarf}) we find that
\be
A=\mu,~~B=\lambda,
\label{AB4}
\ee
and that the momentum $k_y$ is no longer continuous but is quantized  as
\be
k_y^2+\mu^2-(\mu-n)^2=0,~~~n=0,1,2....<[\mu-1].
\ee
The above relation can be satisfied if $k_y=n=0$. Thus in this case the zero energy state is non-degenrate. 
 Here we must take $\psi_A=\psi_B=\psi_-$, where $\psi_-(x)=\phi_0(x)$  in Eq.(\ref{phi_4}) with $n=0$ and $A, B$ given in (\ref{AB4}).

It is interesting to note that if we put $\lam=0$ in (\ref{ex4}) we obtain the following effective potentials
\be\label{ex3-1}
U_\mp(x)=\mu^2-\mu(\mu+1){\rm sech}^2x.
\ee
As in the previous examples the potentials in (\ref{ex3-1}) belong to the shape invariant category and are exactly solvable. The eigenvalues and the corresponding solutions can be obtained from (\ref{scarfsol}) by appropriate change of parameters.

\vspace{.5cm}

\subsection{Example 3}
Here we shall consider a potential of the form
\be
U(x)=ib~\sin(2x),~~b:\,{\rm real}.\label{ex3}
\ee
which leads to the following effective potentials
\be\label{pe1}
U_\pm(x)=b^2\sin^2(2x)\mp 2 b\cos(2x).
\ee
These potentials are of the form
\be\label{pe2}
V(x)=b^2\sin^2(2x)\mp 2ab\cos(2x).
\ee
The potentials in Eq.(\ref{pe2}) are quasi exactly solvable with $a$ band edges of period $\pi$ (if $a$ is odd) ($2\pi$ if $a$ is even) exactly known \cite{ks,R,KM}. Thus comparing the potentials in (\ref{pe1}) and (\ref{pe2}), one concludes that the Dirac equation with the complex potential (\ref{ex3}) has a zero energy state ($\eps=k_y=0$)  with one ($a=1$) 
band edge of period $\pi$.

\vspace{.5cm}

\subsection{ Example 4}

It may be noted that in the previous examples non hermiticity in the model was introduced via a complex coupling constant. However non hermiticity can also be introduced in a different way, namely via a complex coordinate translation. There are a host of examples \cite{khare} which can be treated in this way. As an example, a potential of this class can be taken as
\be
U(x)=-\lambda\, {\rm sech}\, (x-i\mu),~~~~\mu:~ {\rm real}.
\ee
Then from (\ref{upm}) we find
\be\label{sec}
U_\pm(x)=-\lambda^2 {\rm sech}\,^2(x-i\mu)\pm i\lambda {\rm sech}\,(x-i\mu) \tanh\,(x-i\mu)+k_y^2.
\label{U4}
\ee
It may be noted that the zero energy states of the potential (\ref{sec}) for $\mu=0$ have been investigated in a number of papers \cite{zero1,zero2,zero3}. Quasi exact solvability of this potential has also been studied \cite{hartmann}. As in example 2, these potentials can be identified with complexified or more precisely $\cal{PT}$ symmetrized Scarf II potential in (\ref{scarf}).
Comparing $U_+(x)$ with (\ref{scarf}) we get
\be
A=\lam -\frac12,~~ B=\frac12,\n\\
k_y^2=\left(n-\lambda+\frac12\right)^2.
\ee
As in the last example, here $k_y$ is quantized, and  the zero energy state is degenerate with degeneracy $[\lambda-3/2]$. 

\vspace{.5cm}
\section{Dirac equation with Lorentz scalar potential}

In this section we shall briefly discuss Dirac equation with complex potential other than the scalar potential considered in Eq.\,(\ref{dirac1}). 
These other potentials are the Lorentz scalar potential and the vector potential.  

Bound states in massless Dirac equation with complex vector potential has recently been considered  Ref. \refcite{pr}.    So here we will only discuss the Lorentz scalar case.

A $(2+1)$-dimensional Dirac Hamiltonian with a Lorentz scalar potential $U(x)$ has the form (we set  $c=1$)
\begin{equation}
H=\left(\sigma_{x}\hat{p}_x+\sigma_y\hat{p}_y\right)+ \sigma_z U(x)\,.\label{H2}
\end{equation}
 The potential $U(x)$ is assumed to depend only
on $x$.  The corresponding Dirac equation can be cast into a pair of decoupled Sch\"odinger equations.  However, unlike the situation considered in Sect.~3,
the effective potentials for the two components are not energy-dependent.
Exact and quasi-exact solvability of this Dirac equation has been studied in Ref.\refcite{ho}.

One would like to see the potential $U(x)$ could be complexified yet giving a Dirac equation with only real energy eigenvalues.
First let us recapitulate the way 
the Dirac equation with Hamiltonian (\ref{H2}) is reduced to a pair of decoupled Sch\"odinger equations \cite{ho}.  
As $U(x)$ depends on $x$ only, the wave function can be written as
\be
\psi = e^{ik_yy} \left(\begin{array}{c} f_-(x)\\ f_+(x)
\end{array}\right)~,
\label{psi}
\ee
 where $k_y$ is a real constant, and $f_\pm$ are real functions of
 $x$.
 The Dirac equation becomes
\be
\left( \begin{array}{cc} U(x) & p_x-ik_y\\ p_x + ik_y & -U(x)
\end{array}\right)~\left(\begin{array}{c} f_-(x)\\ f_+(x)
\end{array}\right)=E\left(\begin{array}{c} f_-(x)\\ f_+(x)
\end{array}\right)~.
\ee 
Next we
 transform the wave function by a unitary matrix $T$:
\be
\left(\begin{array}{c} f_-(x)\\ f_+(x)
\end{array}\right)\to \left(\begin{array}{c} i\psi_-(x)\\ \psi_+(x)
\end{array}\right)\equiv T^\dagger~\left(\begin{array}{c} f_-(x)\\ f_+(x)
\end{array}\right)~,
~~~T= \frac{1}{\sqrt{2}}\left( \begin{array}{cc} 1 & i\\ i & 1
\end{array}\right)~.
\ee
 Then the Hamiltonian transforms as
\be
H\to T^\dagger H T= \left( \begin{array}{cc} k_y &
i\left(-\frac{d}{dx}+U(x)\right)\\
-i\left(-\frac{d}{dx}+U(x)\right) & -k_y
\end{array}\right)~,
\ee
 and the Dirac equation becomes
\be
\left(\frac{d}{dx} + U(x)\right)\psi_- &=& \left(E+ k_y\right)
\psi_+~,\\ \left(-\frac{d}{dx} +  U(x)\right)\psi_+ &=& \left(E-
k_y\right) \psi_-~, \label{susy-2} \ee 
or
\be
\left(-\frac{d^2}{dr^2} + U^2 \mp U^\prime\right)\psi_\mp =
\epsilon \psi_\mp\,,~~\epsilon=E^2-k_y^2. \label{susy-1} \ee
which is exactly in the
same form as Eq.~(\ref{decoupled}).  
Such form is the factorized, or the supersymmetric form discussed in Ref.\,\refcite{khare}, with $U(x)$ playing the role of the so-called superpotential.

By fitting $U(x)$ in  Eq.\,(\ref{decoupled}) with the superpotential listed Table 4.1 of Ref.\,\refcite{khare}, one finds that there are four cases for which $U(x)$ can be complexified wtth real energies.
These cases are:
\begin{enumerate}
\item[a)]~ {\rm Scar I}: $U(x)=A\,\tan\,x-(B+iC)\, \sec\,x,~~x\in [-\frac{\pi}{2},\frac{\pi}{2}],\\
~~~~~~~~~~~~~~~~\epsilon=(A+n)^2-A^2,~~~ n=0,1,\ldots\, $;
\item[b)]~ {\rm Scarf II }: $U(x)=A\,\tanh\,x+(B+iC)\, {\rm sech}\,x,, ~~x\in (-\infty,\infty),\\
~~~~~~~~~~~~~~~~\epsilon=A^2-(A-n)^2, ~~~n=0,1,\ldots, [A]\,$;
\item[c)]~ {\rm Morse}: $U(x)=A-(B+iC)e^{-x},~~x\in (-\infty,\infty),\\
~~~~~~~~~~~~~~~~\epsilon=A^2-(A-n)^2, ~~~n=0,1,\ldots, [A]\, $;
\item[d)]~ {\rm P\"oschl-Teller}: $U(x)=A\,{\rm coth}\,x-(B+iC)\, {\rm cosech}\, x, ~~x\in [0,\pi],\\
~~~~~~~~~~~~~~~~\epsilon=A^2-(A-n)^2, ~~~n=0,1,\ldots, [A]\,$.
\end{enumerate}
Here the parameters $A, B$ and $C$ are real numbers.

The main characteristics of these four cases is that the energy eigenvalues depend only on one of the parameters, i.e., $A$,  in $U(x)$.  This allows one to complexify the other parameters, namely, $B$ to $B+iC$. The corresponding wave functions can be obtained from those given in Table 4.1 of Ref.\,\refcite{khare} by the corresponding complexification.

\section{Summary}

In this work we have considered $(2+1)$-dimensional $\cal{PT}$-symmetric Dirac equation with a number of complex scalar potentials, either exactly or quasi-exactly solvable. 
In all the cases the solvable eigenstates were determined exactly and the energy eigenvalues were found to be real. Also the coupled equations 
of the two components of the Dirac spinor were shown to be  reducible to a pair of decoupled Schr\"odinger-like  equation, 
but with energy-dependent potential. Thus these equations furnish examples of exactly solvable {\it energy dependent} potentials.   

We have also briefly outlined exactly solvable Dirac equation with complex Lorentz scalar potentials.

\vskip 2truecm

\section*{Acknowledgments}

The work is supported in part by the Ministry of Science and Technology (MOST)
of the Republic of China under Grant NSC-102-2112-M-032-003-MY3 (CLH). 
PR wishes to
thank the R.O.C.'s National Center for Theoretical Sciences (North Branch)  and
National Taiwan University for supporting a visit during which part of the work was done.



\begin{thebibliography}{0}

\bibitem{b1} C.M. Bender and S. Boettcher, {\it Phys. Rev. Lett.} {\bf 80}, 5243 (1998).

\bibitem{mostafa} A. Mostafazadeh, {\it J. Math. Phys.} {\bf 43}, 205 (2002); {\it  ibid} {\bf 43}, 2814 (2002).

\bibitem{rel1} A. Sinha and P. Roy, {\it Mod. Phys. Lett.} {\bf A20}, 2377 (2005).

\bibitem{rel2} F. Cannata and A. Ventura, {\it Phys. Lett. } {\bf A372}, 941 (2008).

\bibitem{rel3} V.G.C.S. dos Santos, A. de Souza Dutra and M.B. Hott, {\it Phys. Lett.} {\bf A373}, 3401 (2009).

\bibitem{rel4} F. Cannata and A. Ventura, {\it J. Phys.} {\bf A43}, 075305 (2010).

\bibitem{rel5} B.P. Mandal and S. Gupta, {\it Mod. Phys. Lett.} {\bf A25}, 1723 (2010).

\bibitem{longhi} S. Longhi, {\it Phys. Rev. Lett.} {\bf 105}, 013903 (2010).

\bibitem{zafar} Z. Ahmed, {\it Phys. Lett.} {\bf A282}, 343 (2001).

\bibitem{al} A.D. Alhaidary, {\it Phys. Lett. }{\bf A377}, 2003 (2013).

\bibitem{GGPS} J. Garcia-Martinez, J. Garcia-Ravelo, J.J. Pena and A. Schulze-Halberg, {\it Phys. Lett.} {\bf A373}, 3619 (2009).

\bibitem{YLL} R. Yekken, M. Lassaut and R.J. Lombard, {\it Ann. Phys.} {\bf 338}, 195 (2013).

\bibitem{dirac} E. Merzbacher, {\it Quamtum Mchanics}, 3rd ed. (John Wiley, 1998).

\bibitem{khare} F. Cooper, A. Khare and U. Sukhatme, {\it Supersymmetry in Quantum Mechanics} (World Scientific,  2001). 

\bibitem{ks} A. Khare and U. Sukhatme, {\it J. Phys. }{\bf A37}, 10037 (2004).

\bibitem{R} M. Razavy, {\it Am. J. Phys.} {\bf 48}, 285 (1980).

\bibitem{KM} A. Khare and B.P. Mandal, {\it J. Math. Phys.} {\bf 39}, 3476 (1998). 

\bibitem{zero1} R.R. Hartmann, N.J. Robinson, and M.E. Portnoi, {\it Phys. Rev.} {\bf B 81}, 245431 (2010).

\bibitem{zero2} D.A. Stone, C.A. Downing, and M.E. Portnoi, {\it Phys. Rev.} {\bf  B 86}, 075464 (2012).

\bibitem{zero3} R.R. Hartmann, I.A. Shelykh, and M.E. Portnoi, {\it Phys. Rev.}{\bf B 84}, 035437 (2011).

\bibitem{hartmann} R.R. Hartmann and M.E. Portnoi, {\it Phys. Rev.} {\bf A89}, 012101 (2014).

\bibitem{pr}
O. Panella and P. Roy, {\it Symmetry} {\bf 6}, 103 (2014).

\bibitem{ho}
C.-L. Ho, {\it Annals of Physics} {\bf 321}, 2170  (2006), Sect. 4.1.


\end{thebibliography}
\end{document}